\documentclass[preprint2]{aastex}

\usepackage{amstex,psfig,epsfig,natbib}

\newcommand{\AIPS}{{$\cal AIPS\/$}}

\begin{document}

\title{VLA 1.4GHz observations of the GOODS-North Field: 
Data Reduction and Analysis}

\author{Glenn E.\ Morrison,\altaffilmark{1,2} Frazer N.\
Owen,\altaffilmark{3} Mark Dickinson,\altaffilmark{4} R.\,J.\
Ivison\altaffilmark{5,6} and Edo Ibar\altaffilmark{5}}

\medskip

\altaffiltext{1}{Institute for Astronomy, University of Hawaii, 
Manoa, Hawaii 96822}

\altaffiltext{2}{Canada-France-Hawaii Telescope Corp.,
Kamuela, Hawaii 96743}

\altaffiltext{3}{National Radio Astronomy Observatory, P.\ O.\ Box O, Socorro,
NM 87801}

\altaffiltext{4}{National Optical Astronomy Observatory, P.\ O.\ Box 26732,
Tucson, Arizona 85726}

\altaffiltext{5}{UK Astronomy Technology Centre, Royal Observatory, 
Blackford Hill, Edinburgh EH9 3HJ, UK}

\altaffiltext{6}{Institute for Astronomy, University of Edinburgh,
Blackford Hill, Edinburgh EH9 3HJ, UK}

\begin{abstract}
We describe deep, new, wide-field radio continuum observations of the
Great Observatories Origins Deep Survey -- North (GOODS-N) field.  The
resulting map has a synthesized beamsize of $\sim$1.7\arcsec\ and an
r.m.s.\ noise level of $\sim$3.9\,$\mu$Jy\,beam$^{-1}$ near its center
and $\sim$8\,$\mu$Jy\,beam$^{-1}$ at 15\arcmin\, from phase center.
We have cataloged 1,230 discrete radio emitters, within a
40\arcmin\,$\times$\,40\arcmin region, above a 5-$\sigma$ detection
threshold of $\sim$20\,$\mu$Jy at the field center. New techniques,
pioneered by \cite{owe08}, have enabled us to achieve a dynamic range
of 6800:1 in a field that has significantly strong confusing sources.
We compare the 1.4-GHz (20-cm) source counts with those from other
published radio surveys. Our differential counts are nearly Euclidean
below 100\,$\mu$Jy with a median source diameter of
$\sim$1.2\arcsec. This adds to the evidence presented by \cite{owe08}
that the natural confusion limit may lie near 1\,$\mu$Jy.  If the
Euclidean slope of the counts continues down to the natural confusion
limit as an extrapolation of our log N - log S, this indicates that
the cutoff must be fairly sharp below 1 $\mu$Jy else the cosmic
microwave background temperature would increase above 2.7K at 1.4 GHz.

\end{abstract}


\section{Introduction}

The GOODS-N field \citep{dic03,gia04} covers $\approx160$\,arcmin$^{2}$
centered on the Hubble Deep Field North \citep{wil96} and is
unrivaled in terms of its ancillary data. These include extremely deep
{\em Chandra}, {\em Hubble Space Telescope} and {\em Spitzer}
observations, deep $UBVRIJHK$ ground-based imaging and $\sim$3,500
spectroscopic redshifts from 8--10-m telescopes. Previous radio
observations of this region, however, fell  short of complementing
this unique dataset.

\begin{deluxetable}{cccc}
\tablecolumns{4}
\tablewidth{0pt}
\tablecaption{GOODS-North VLA observing log\label{TABLE1}}
\tablenum{1}
\pagestyle{empty}
\tablehead{
\colhead{Observing dates}& 
\colhead{Integration time (h)}&
\colhead{NRAO project code}&
\colhead{Configuration}}
\startdata

Nov-Dec 1996 & 42  & AR368& A\\
Feb 2005     & 28  & AM825& B\\
Aug 2005     & 7   & AM825& C\\
Dec 2005     & 2   & AM825& D\\
Feb-Apr 2006 & 86  & AM857& A\\
\enddata
\end{deluxetable}

Radio emission is a relatively unbiased tracer of star formation and
can probe heavily obscured active galactic nuclei (AGN) -- objects
that are missed by even the deepest X-ray surveys. Radio observations
thus allow us to fully exploit the wealth of data taken at
X-ray--through--millimeter wavelengths, providing a unique
extinction-free probe of galaxy growth and evolution through the
detection of starbursts and AGN.

The recent imaging of \cite{owe08} (OM08) ($\sigma =
2.7\,\mu$Jy\,beam$^{-1}$ at 1.4\,GHz) have shown that the techniques
exist to make radio images that approach the theoretical noise
limit. To this end, we have obtained new, deep radio imaging of the
GOODS-N field.

\begin{figure*}[!ht]
\begin{center}
\epsscale{2.0}\plotone{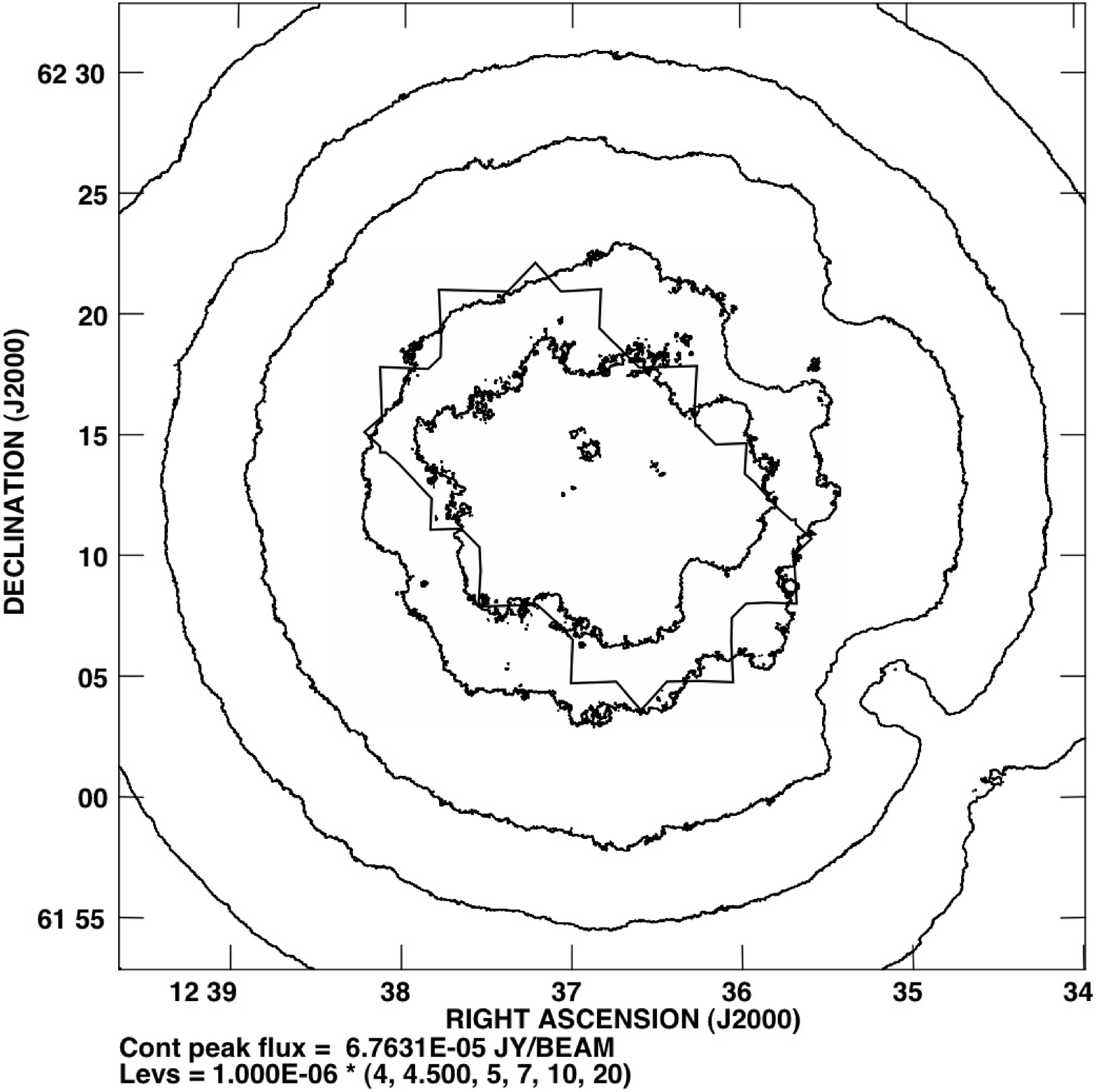}
\caption{Contours of constant r.m.s.\ noise after correction for the
shape of the primary beam. The GOODS-N area covered by {\em HST}/ACS
is outlined near the field center, with an average sensitivity of
$\sim$3.9\,$\mu$Jy\,beam$^{-1}$.  At a radius of $\sim$15\arcmin\
(roughly the half-power point of the primary beam) the r.m.s.\
sensitivity drops to $\sim$7\,$\mu$Jy\,beam$^{-1}$. The strong inward
deflection of the 7-$\mu$Jy and 10-$\mu$Jy contours is due to the
265-mJy radio source discussed in \S\ref{DynRange}.}\label{fig1}
\end{center}
\end{figure*}

While GOODS-N was selected to be free from bright sources at optical
wavelengths, the field contains several very bright radio sources
which place severe limitations on the dynamic range that can be
obtained and hence the ultimate sensitivity of the radio map. Before
new techniques were developed to deal with this issue, only
moderately-deep radio imaging was possible \citep{ric00}.

The earliest VLA data in the GOODS-N field were reprocessed using new
techniques by \cite{big06} and \cite{mor08}, achieving a noise level
of 5.7--5.8\,$\mu$Jy\,beam$^{-1}$ -- a 23-25\% improvement on the
original map by \citet{ric00}, and close to the theoretical noise
limit when one considers the increase in system temperature at the low
elevations permitted during the original observations.

While the reduction of \cite{big06} provided improved access to the
$\mu$Jy radio population, even deeper radio imaging is required to
properly complement the extremely deep GOODS {\em Spitzer}
mid-infrared data, as well as forthcoming deep observations at
far-infrared and submillimeter wavelengths from {\em Herschel},
SCUBA-2, the Large Millimeter Telescope, and other facilities. To this
end, we have added 123\,hr to the existing data. The reduced full
resolution image (beam=1.7\arcsec) and its rms map are available
online\footnote{http:/www.ifa.hawaii.edu/$\sim$morrison/GOODSN and at
the NASA/IPAC Infrared Science Archive (IRSA)
http://irsa.ipac.caltech.edu/ as an ancillary data product associated
with the GOODS Spitzer Legacy survey}.

The paper is laid out as follows: in \S2 we describe the observations
and the reduction of the data. In \S3 we discuss the cataloging of
radio emitters. \S4 contains the results and some discussion of the
catalog. We present our conclusions in \S5.

\section{Observations, Reductions, and Cataloging}

In 1996 November, \cite{ric00} observed a region centered at
12:36:49.4, +62:12:58 (J2000) for a total of 50\,hr at 1.4\,GHz using
the National Radio Astronomy Observatory's (NRAO's) VLA\footnote{The
NRAO is operated by Associated Universities, Inc., under a cooperative
agreement with the National Science Foundation.} in its A
configuration. Of this, only 42\,hr was considered usable by
\citet{ric00}. Adopting the same position and frequency, we obtained
28\,hr of data in the VLA's B configuration in February--April 2005,
7\,hr in C configuration in August 2005, and 2\,hr in D configuration
in December 2005, and 86\,hr in A configuration in 2006
February--April (see Table~\ref{TABLE1}) -- for a useful combined
total of 165\,hr. Observations were done at night to avoid solar
interference. We followed the 1:4 scaling of visibility data between
the arrays described by OM08. This empirically derived scaling
relation provides for more uniform weighting of $uv$ data.

In most regards the new observations were taken using the same
parameters as those used in 1996. However, the integration time was
changed from 3.33\,s to 5\,s because of difficulties experienced by
the correlator with the shorter integration time \citep{ric00}. The
data were all obtained using spectral-line mode 4, which yields
7\,$\times$\,3.125-MHz channels in each of two intermediate
frequencies (IFs), centered at 1,365 and 1,435\,MHz, in each of two
circular polarizations. The channel width and integration time were
compromises chosen to maximize reliability and sensitivity while
minimizing bandwidth (radial) and time (tangential) smearing effects,
respectively. The upcoming EVLA correlator, WIDAR, will offer much
shorter integration times, narrower channels, and greater overall
bandwidth.

\subsection{Calibration and editing}

\AIPS\ was used to reduce and analyze all the radio data.  
The first step was the calculation and corrections of the spectral
bandpass shape.  This was done in the following manner using the task
{\sc BPASS}.  The bright point-source phase calibrator was {\sc split}
from the raw database and phase self-calibration was applied. The
self-calibrated data were then used to calculate a bandpass correction
and flatten the spectral response across the band for the uncalibrated
multi-source database.  Standard flux density calibration was applied
next, using the Baars flux-density scale with 3C\,286 \citep{baa77} as
the calibrator. The antenna-based weights for each 5-s integration
were also calibrated. The $uv$ data for the target field were {\sc
split} from the database and clipped using the \AIPS\ task {\sc clip}
at a level well above the total flux density found in the field to
remove any interference prior to the self-calibration process. Only
minor interference was encountered during the observations.

\subsubsection{UVFIX - Calculating $u, v,$ and $w$}

In verifying the astrometry of the A-array data from 1996 and then
again in 2006, we found a rotation between the two astrometric
frames.  The rotational offset was about 1\arcsec\ at a radial distance of
20\arcmin\ from the phase center, and is likely the cause of the
optical--radio offsets reported regularly over the last decade
\citep[e.g.][]{ivi02}.  The offset problem was traced back to the VLA
on-line system (the so-called MODCOMPS) and its calculated $u, v,$ and
$w$ terms. This problem was corrected by using the AIPS task {\sc
UVFIX}. This task computes values of $u, v,$ and $w$ using the time
index and baseline information recorded during the observations.  The
\AIPS\ task {\sc UVFIX} was used on all the $uv$ datasets after the
initial bandpass and flux calibration.

\subsection{Imaging and self-calibration}

Bright sources and their sidelobe patterns affect the noise structure
and the dynamic range of the final reduced maps if they are not
cleaned. Such sources were located using D-configuration data. We
imaged to a radius of 8\,degrees using a pixel size of 30\arcsec. The
GOODS-N field was chosen to be free from bright objects at most
wavelengths, but not the radio waveband. As a result, bright radio
sources limit the dynamic range. We imaged sources as far as
1.5\,degrees from the phase center -- a total of 56 facets, selected
to cover distant, bright sources, plus another five facets within the
primary beam to help with the {\sc PEELR} process which is fully
discussed in OM08.

Each facet is constructed from the Fourier transform of the data,
phase shifted to its tangent point on the celestial sphere. In all, 98
facets were used. This technique is known as polyhedral imaging 
(also known as faceting) and is discussed at length by \cite{cor92}.

One night of A-configuration data was used to lay out 37 facets to
cover the primary beam at 1.4\,GHz. Each of the 37 facets comprised
1024$^2$ pixels, each 0.5\arcsec\, pixels. The other 56 facets were
512$^2$-pixels in size, again with 0.5$^2$\arcsec pixels, centered on
the bright, outer sources.

Initial cleaning of the fields with {\sc IMAGR} revealed faint sources
within the 98 facets and tight clean boxes were placed around each
one, thus limiting cleaning to real features. Maps were cleaned down
to the 1-$\sigma$ noise level to create clean components for use in
the self-calibration process. Clean boxes are a vital aspect of this
process. They avoid the possibility of cleaning into the noise, which
can change the noise structure of a map and result in so-called `clean
bias'. Failure to use clean boxes can artificially reduce the noise.

This one night of $uv$ data were first self-calibrated in phase only,
via the \AIPS\ task {\sc CALIB}, and a new map was made.
Self-calibration in amplitude and phase followed. The clean components
from this one night were used to construct the fiducial model for the
remaining 13 nights of new A-configuration data (AM857 -- NRAO project
number) as well as the seven nights of 1996 (AR368) A-configuration
data. For the four nights of new B-configuration data (AM825), one of
the B-configuration datasets were used to provide the clean model for
self-calibration. The C- and D-configuration data (AM825) were
bootstrapped to the clean model in a similar fashion.

Editing of the $uv$ data was done before and after the
self-calibration process using the \AIPS\ task {\sc TVFLG}.  After
each episode of self-calibration, the clean component model was
removed from the $uv$ database using {\sc UVSUB}. {\sc UVPLT} was then
used to determine the residual amplitude distribution. Any high
residual points were deleted with {\sc clip}, except those on the
shortest baselines.

The 165\,hr of data resulted in a very large database so we reduced
the data volume by exploiting the fact that the A- and B-configuration
(programs AM825 and AM857) observations were taken at the same hour
angles ($\pm$3.5\,hr of transit) over many nights. The smaller hour
angle interval helped mitigate the higher system temperature that
occurred at lower dish elevations. This is due to the fact that the
antenna feeds see the ground at low elevations.  The AR368 observing
tracks covered $\pm$7.5\,hr of transit yielding extended periods of
low-elevation observations. The antenna weights were calibrated as
part of the standard calibration process minimizing the impact of the
more noisy low-elevation data. Finally, {\sc STUFFR} is used to
average data within a small range of u,v (so there is no effect on the
image), which was weighted properly. In this way the number of data
points used for the imaging were reduced by an
order-of-magnitude. This is fully described in OM08.

The five separate datasets were combined into one large $uv$ dataset
using {\sc DBCON}. This was then divided up by hour angle, by Stokes
parameter (LL and RR, separately) and by IF, thus yielding eight
separate datasets. This last step minimizes the problem of `beam
squint' wherein the two circular polarizations have different pointing
centers due to the slight off-axis location of the feeds. This,
combined with the alt-az telescope mount, causes the effective gain
for an off-axis source to vary as a function of time. Moreover, the
two IFs are at different frequencies and so yield a different primary
beam shapes hence correspondingly different gains for off-axis
sources.


\begin{center}
\begin{deluxetable} {crrrccccccc}
\tablecaption{Sample table of radio sources\label{TABLE2}}
\tabletypesize{\scriptsize}
\tablewidth{0pt}
\tablenum{2}
\pagestyle{empty}
\tablehead{
\colhead{Number}&
\colhead{R.A.\ (J2000)}&
\colhead{Dec.}&
\colhead{SNR}&
\colhead{$S_p$}&
\colhead{$S_i$}&
\colhead{}&
\colhead{Size}&
\colhead{}&
\colhead{Upper}&
\colhead{Beam}\\
\colhead{}&
\colhead{hms $\pm$ e(s)}&
\colhead{d \,\arcmin\, \arcsec $\pm$ e(\arcsec)}&
\colhead{}&
\colhead{($\mu$Jy\,bm$^{-1}$)}&
\colhead{($\mu$Jy)}&
\colhead{Maj}&
\colhead{Min}&
\colhead{P.A.}&
\colhead{(\arcsec)}&
\colhead{(\arcsec)}}

\startdata

    5 &12 34  3.51 0.061 &62 14 20.7 0.03 &  27.6&    191.6$\pm$  6.9&     795.9$\pm$   44.9&   2.6&  0.7&  95&  0.0& 1.7 \\
    9 &12 34  8.79 0.164 &62  9 20.5 0.10 &   8.0&     51.6$\pm$  6.5&     142.4$\pm$   19.5&   0.0&  0.0&   0&  3.8& 1.7 \\
   12 &12 34  9.23 0.098 &61 56 45.5 0.09 &  44.3&    397.0$\pm$  8.9&    2767.1$\pm$  171.5&   7.2&  0.0& 129&  0.0& 6.0 \\
   13 &12 34  9.87 0.238 &62  3 57.6 0.12 &   7.3&     57.3$\pm$  7.9&     208.6$\pm$   30.4&   2.8&  0.0&  89&  0.0& 1.7 \\
   14 &12 34 10.61 0.085 &62  6 15.8 0.05 &  18.6&    132.1$\pm$  7.1&     564.4$\pm$   43.5&   2.4&  0.8&  67&  0.0& 1.7 \\
   18 &12 34 10.78 0.083 &62  2 54.8 0.05 &  19.2&    173.4$\pm$  9.0&     848.0$\pm$   63.7&   2.6&  0.5&  63&  0.0& 1.7 \\
   22 &12 34 11.16 0.029 &62 16 17.3 0.02 & 239.1&   1674.0$\pm$  7.0&    7741.0$\pm$  236.7&   8.5&  1.6& 126&  0.0& 6.0 \\
   23 &12 34 11.74 0.003 &61 58 32.5 0.00 & 438.5&   4385.0$\pm$ 10.0&   27684.0$\pm$  834.7&   1.7&  0.6&  52&  0.0& 1.7 \\
   35 &12 34 16.09 0.071 &62  7 22.4 0.05 &  22.8&    127.5$\pm$  5.6&     346.2$\pm$   18.2&   0.0&  0.0&   0&  3.7& 3.0 \\
   36 &12 34 16.17 0.157 &62 25 58.8 0.13 &   8.6&     66.1$\pm$  7.7&     284.2$\pm$   34.6&   0.0&  0.0&   0&  3.6& 1.7 \\

\enddata
\end{deluxetable}
\end{center}


After self-calibration, some artifacts associated with two bright
sources (within the 37 facets that map out the primary beam)
remained. A local self-calibration technique in \AIPS\ called, {\sc
PEELR was used to remove these sidelobes. The method is described as
follows.

In brief, {\sc PEELR} subtracts the clean component (CC) model for all
facets from the $uv$ data, except for the facet containing the source
to be `peeled'. Then {\sc PEELR} self-calibrates using this
information (flux) in this facet, writes a new calibration
table. Next, {\sc PEELR} removes the CCs for the chosen field, thereby
removing the bright source. The special calibration for this one field
aids in the sidelobe correction process. {\sc PEELR} then goes back to
the original calibration tables and adds the CCs back to the $uv$
dataset. Having the $uv$ data in eight separate files allows for a
more accurate CC model, enabling more accurate removal of these bright
sources. }

Full details on this important process can be found in OM08.

\subsubsection{Final signal and noise images}

The final maps were constructed by combining the 37 central facets
made with each of the eight $uv$ datasets, weighting them by
$1/\sigma^2$, using {\sc FLATN}.  The r.m.s., before correcting for
the shape of the primary beam, is $\sim$3.9\,$\mu$Jy over a region of
100$^2$ pixels. The synthesized beam size is
1.7\arcsec\,$\times$\,1.6\arcsec\ with a position angle of
$-$5\,degrees.  The final step was to run \AIPS\ task {\sc MWFLT}
which low-pass filters the image with a 101$^2$-pixel kernel to yield
a more uniform background.

The \AIPS\ task, {\sc RMSD}, was used to construct a noise image,
calculating a histogram based r.m.s.\ for each pixel using the
surrounding pixels within a radius of 100 pixels.  A multi-iteration
process rejects pixels outside of the $\pm3\sigma$ range, thereby
offering a more robust estimate of the noise.  Fig.~\ref{fig1} shows
contours of constant noise over the central region, after correcting
for the primary beam response.

\section{Cataloging}

\subsection{ Angular size effects}

The angular size of discrete sources in the image are broadened
by three effects: (1) the finite bandwidth of each channel where

\begin{equation}
{\rm BWS} \sim \frac{\Delta \nu}{\nu} \times (r),
\end{equation}

\noindent where $\Delta \nu$ is the fractional bandwidth
(i.e. 3.125\,MHz) and $r$ is the distance from the phase center; (2)
is the finite data sampling rate (estimated at a few percent); (3) and
the true angular size of the source. Thus, in order to detect sources
above 4.5-sigma, images at 3\arcsec and 6\arcsec are also needed to
detect extended sources--whether truly extended or experimentally
extended.  OM08 showed that 3\arcsec\ and 6\arcsec\ maps are a useful
complement to the full-resolution images to recover the full flux of
extended sources. The 3\arcsec\ and 6\arcsec\ images were made using
the \AIPS\ task, {\sc CONVL}.

\subsection{Source extraction and cataloging}

The \AIPS\ task, {\sc sad}, (`search and destroy') was used to
generate an initial source catalog. For each resolution, {\sc sad} was
used in `signal-to-noise ratio (SNR) mode' to search for peaks more
than 4.5$\times$ the local noise and to correct for radial smearing
and primary beam attenuation. This mode uses the noise maps created by
{\sc RMSD}.  The final catalog was clipped at $\geq$5$\sigma$.

Residual maps created by {\sc sad} were searched by eye to find
sources missed by the automatic procedure. We made SNR maps from the
residual maps, searching for peaks above four. This lower SNR limit
was picked because BWS reduces the peak flux of a source. Next, the
properties of these sources were measured using \AIPS\ task, {\sc
JMFIT}. Sources with SNR greater than 4.5 were retained.  Any such
sources were added to the catalog for the appropriate resolution.

{\sc SAD} sources with SNR $\leq$ 5.5 were re-measured by hand using
{\sc JMFIT} -- the same task used by {\sc sad} to determine the source
properties. At this low flux limit {\sc sad}'s automatic detection
routine does not always choose the optimal area for analysis.  Those
pixels within a defined radius (100 pixels) were used to calculate the
local noise for each source.

Following OM08, resolved sources were classified on the basis of the
best-Gaussian-fit major axis. If the lower limit for this parameter
was greater than zero, the source is classed as resolved and the
integrated flux was used for the total flux. If lower limit for the
major axis was equal to zero (i.e., unresolved) then the peak flux was
adopted as the total flux for the source.

There are several extended sources within the central field, as shown
in Fig.~\ref{kntr2}. {\sc TVSTAT} was used to measure the total flux
density of extended sources. The SNRs of extended sources were
determined as follows: {\sc IMEAN} was used to determine the brightest
peak and its position was then examined in the noise map; the ratio
of these values was adopted as SNR.

Results were then collated following the prescription given in OM08.
An additional noise term of 3\,\% was included for each source to
account for calibration errors and uncertainties in the primary beam
correction.

\section{Results and discussion}

\subsection{Radio catalog}

A sample of the radio catalog is given in Table~\ref{TABLE2}. The
complete catalog is available electronically. Column 1 is the source
number: source numbers below 10,000 relate to those sources found by
{\sc sad} while those above 10,000 were those found and investigated
by hand. Columns 2 and 3 contain the right ascension (R.A.) and the
declination (Dec.) in J2000, with the positional error. Column 4
represents the SNR using the ratio of the observed peak flux density
and the local noise. In columns 5 we list the peak flux density and
its error, in $\mu$Jy\,beam$^{-1}$.  Columns 6 contain the total,
primary-beam-corrected flux density and uncertainty, in $\mu$Jy. The
best-fit deconvolved size (in arcsec) is given in column 7-9. Using
our criteria for resolved sources, if a resolved 2-dimensional
Gaussian size was the best fit then we give the resulting major and
minor {\sc fwhm} size (in arcsec) along with the position angle
(P.A.). If this was not the case then in column 10 we report the major
axis upper limit, as estimated by {\sc JMFIT} or {\sc SAD}.  If
the source is resolved then we report 0.0\arcsec\, for the upper limit
size. As noted above, very extended sources, as shown in
Fig.~\ref{kntr2}-\ref{kntr4}, had their positions, sizes, fluxes and
errors measured interactively using {\sc IMVAL} and {\sc TVSTAT}. The
final column reports the resolution of the map that was used for the
measurements for that particular source.

\begin{deluxetable}{cccccccccl}
\tablecolumns{8}
\tablewidth{0pt}
\tablecaption{Source size summary\label{TABLE3}}
\tablenum{3}
\pagestyle{empty}
\tablehead{
\colhead{Radius}& 
\colhead{Minimum}&
\colhead{Maximum}&
\colhead{Sources}&
\colhead{Upper}&
\colhead{Mean}&
\colhead{Error}&
\colhead{Median}&
\colhead{Resolved}\\
\colhead{(arcmin)}&
\colhead{($\mu$Jy)}&
\colhead{($\mu$Jy)}&
\colhead{}&
\colhead{limit}&
\colhead{(arcsec)}&
\colhead{(arcsec)}&
\colhead{(arcsec)}&
\colhead{(\%)}}
\startdata	
5 & 20 & 30  & 32 & 37& 1.2& 0.1& 1.16& 46\\
5 & 30 & 100 & 49 & 24& 1.8& 0.2& 1.15& 67 \\
10& 100& 300 & 34 & 26& 2.3& 0.4& 1.09& 57\\
20& 300& 1000& 48 & 11& 3.4& 0.6& 1.57& 81\\
\enddata
\end{deluxetable}

\subsection{Angular size distribution}

As a check on the radio angular size distribution in the GOODS-N
field, we have repeated the source size analysis of OM08. The focus of
this analysis is to estimate the angular size of the faint radio
population. As noted in column~1 of Table~\ref{TABLE3}, we have
analyzed the more sensitive inner region ($<$5\arcmin) of the radio
map for fainter sources while for the brighter sources we expanded our
search outward (to 20\arcmin) to improve the statistics. Columns 2 and
3 give the flux range and column 4 lists the  sources within that
size range. Column 5 shows the number of upper limits used with the
measured source size distribution in the Kaplan-Meier
analysis\footnote{IRAF task {\sc kmestimate} and algorithm by
\cite{fei85}} to derive a robust mean size (column 6) and its
associated error (column 7). The median size is reported in column 8
and the percentage of resolved sources used in this analysis is listed
in column 9.

The results in Table~\ref{TABLE3} are consistent with both OM08 and
\cite{mux05}, adding to the evidence that the size distribution of the
radio population does not continue to decrease below 100\,$\mu$Jy and
that the median size of such faint emitters is around 1\arcsec.

\begin{deluxetable}{rrccc}
\tablecolumns{6}
\tablewidth{0pt}
\tablecaption{Differential normalized source counts for GOODS-N\label{TABLE4}}
\tablenum{4}
\pagestyle{empty}
\tablehead{
\colhead{$S_{\rm l}$}& 
\colhead{$S_{\rm h}$}&
\colhead{$S_{\rm ave}$}&
\colhead{No.}&
\colhead{$S^{2.5}dN/dS$}\\
\colhead{($\mu$Jy)}&
\colhead{($\mu$Jy)}&
\colhead{($\mu$Jy)}&
\colhead{\#}&
\colhead{(Jy$^{1.5}$\,sr$^{-1}$)}}
\startdata	
  28.0&        40.0&   34.0&  185&   7.37$\pm$0.70\\
  40.0&        55.0&   46.7&  187&   7.45$\pm$0.89\\
  55.0&        90.0&   71.1&  274&   6.67$\pm$0.70\\
  90.0&       150.0&  113.1&  150&   7.39$\pm$1.31\\
 150.0&       300.0&  199.8&  107&   5.29$\pm$1.11\\
 300.0&      1,500.0&  557.3&   65&   6.02$\pm$1.79\\
1,500.0&     60,000.0& 5,260.5&   17&   5.56$\pm$1.68\\
\enddata
\end{deluxetable}

\subsection{Log $N$--Log $S$}

We have determined the differential normalized sources counts from our
radio catalog using the code and analysis techniques described in
detail by OM08. The results are tabulated in Table~\ref{TABLE4}. The
relative point-source sensitivity as a function of the distance from
the field center for the three resolutions can be seen in figure~4 of
OM08. Their figure~5 also shows the relative sensitivity as a function
of source size. To summarize, multiple effects can make the source
list incomplete, even at rather large SNR. The sensitivity for
resolved sources is thus a function of both beam size and distance
from the phase center. These effects include (i) primary beam
attenuation, (ii) reduction of peak flux due to BWS, (iii) finite
time-average smearing and (iv) resolution bias. Sources with
integrated fluxes near the survey limit, if extended, will have peak
sources below the limit.

As seen in Table~\ref{TABLE3}, the median source size is similar to
that of the synthesized beam ($\sim$1.7\arcsec). To correct for the
effects of this, we adopt the OM08 size distribution, which is based
on the properties of sources at higher SNR.

\subsection{Comparison to previous surveys}

\subsubsection{Dynamic range}\label{DynRange}

As described by \cite{big06}, and seen in their figure~2, the
brightest GOODS-N radio source within the primary beam of the VLA is
near 12:34:52, $+$62:02:35. With a primary-beam-corrected total flux
of 263\,mJy and a peak flux of $\sim$86\,mJy\,beam$^{-1}$, it is
bright sources like this one (see Fig.~\ref{kntr3} upper right) that
can limit the dynamic range in an image. This is because there is a
limit to how well we can clean these bright sources -- a limit
determined by how well we can characterize the bandpass, the pointing,
the antenna-based gains, and many other parameters. As a result of the
imperfect cleaning, residual flux spills into the image from the
sidelobes of bright sources. Peak flux density, in Jy\,beam$^{-1}$, is
therefore a key quantity when determining the suitability of a field
for deep radio imaging, where the `beam$^{-1}$' indicates a dependence
on resolution.

The reduction by \cite{ric00} yielded an image with a dynamic range of
$\sim2,800:1$, with $\sigma=7.5\,\mu$Jy. This was due in part to the
fact that self-calibration was not applied to the $uv$
data. Self-calibration and newly developed techniques allowed
\cite{big06} to create a more sensitive map, with $\sigma=5.8\,\mu$Jy
and a dynamic range of $\sim4,400:1$. Our reduction, which includes a
large quantity of new data (in all VLA configurations) , reaches
$\sigma=3.9\,\mu$Jy\,beam$^{-1}$ and a dynamic range of $\sim6,800:1$.

The $\sim$140-hr VLA observation by \cite{owe08} -- in 1046+59, a
field deliberately chosen to be free of strong radio sources -- has
$\sigma=2.7\,\mu$Jy, which is close to the theoretical noise limit,
yet the dynamic range is only $\sim2,500:1$ because of the paucity of
bright radio emitters.

\subsubsection{Log $N$--log $S$}

\begin{figure*}[!ht]
\begin{center}
\includegraphics[angle=90,width=6.5in]{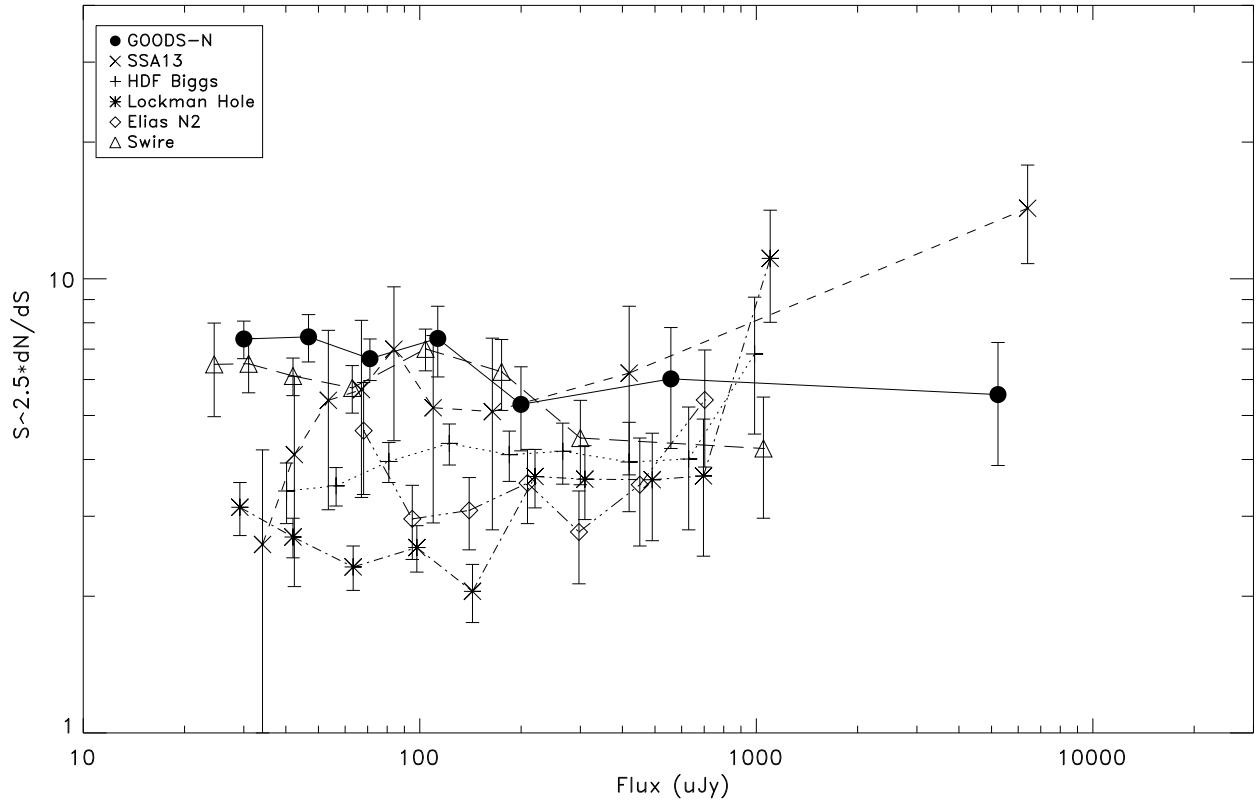}
\caption{Log $N$--log $S$ for corrected differential source
counts below 1\,mJy at 1.4\,GHz. HDF Biggs, Lockman Hole and ELAIS
N2 are from \cite{big06} and SSA\,13 is from \cite{fom06}.
The 1046+59 data are from \cite{owe08}.}
\label{fig2}
\end{center}
\end{figure*}

Fig.~\ref{fig2} shows our differential source counts below 1\,mJy,
compared to other deep, 1.4-GHz continuum surveys. Our source counts
agree with those for the 1046+59 field where the same methodology was
employed for cataloging and for determining log $N$--log $S$: use of
multi-resolution maps to build a master source catalog, and the
adoption of the same source size distribution.

Above 100\,$\mu$Jy our new survey agrees with the counts presented by
\cite{big06}. Below 100\,$\mu$Jy, the counts in the new survey flatten
out while those of \cite{big06} began to turn down. This is
a well-known issue in deep radio surveys. The difference can be
related to a number of complications:

\begin{enumerate}
\item
We believe that one of the main problems arises from the Gaussian fit
not always being the best approach to estimate source
properties. Indeed, the intrinsic shape of the sources gets badly
affected due to bandwidth and time delay smearing. We have found
evidence that flux densities are underestimated systematically after
sources have been convolved to lower resolutions -- flux densities are
larger in the convolved maps. Also, a particular uncertainty comes
from the assumption that in unresolved sources we consider the peak
flux density as an estimate of the total flux density. These problems
clearly suggest the necessity for a more robust source extraction
method in radio observations.  Another, smaller uncertainties coming
from the `clean bias' and `flux boosting' are not expected to be
important due to the tight clean boxes we used, and the high
5-$\sigma$ threshold in the source extraction, respectively.

\item

A knowledge for the intrinsic radio source size distribution is
essential when estimating the number of sources missed by our imaging
approach. \citet{mux05} use high-resolution MERLIN and A-configuration
VLA data to tackle this problem, but they miss a non negligible number
of low-surface-brightness sources, as is indicated clearly by
cross-matching the catalog with lower resolution observations using
the WSRT. We believe the combination of VLA data in its A, B, C, and D
configurations followed by a source extraction performed in convolved
images is essential for detecting these low surface brightness sources
($>$\,4\arcsec). A simple source extraction using SAD in the original
non-convolved image misses a considerable number of sources as we show
in this work. We find that our survey is in good agreement with OM08,
so given by the similar treatment, we have adopted their source size
distribution for number count estimations.

\item

Finally, the source extraction based on a SNR criterion highly depends
on reliable estimates of noise as a function of position in the map
and the efficiency for detection. Monte-Carlo simulations for the
source extraction method (e.g. inserting randomly placed sources in
the map and then doing a random search to extract them) are typically
used for these purposes. Nevertheless, given our detailed source
extraction and our constant examination of residual maps, we believe
that our number counts do not suffer significantly from incompleteness
down to a threshold of 5$\sigma$. This is why we prefer to perform
theoretical corrections, as explained in OM08, rather than a simple
Monte-Carlo approach as was in \cite{iba09}.

\end{enumerate}


\subsubsection{Log $N$--log $S$ Euclidean nature below 100\,$\mu$Jy}

It is expected that the Cosmic microwave background (CMB) temperature
would increase above 2.7K at 1.4 GHz if the counts do not turn over at
some flux density below where we have
reached. \citep[e.g.][]{ger08}. The Log $N$--log $S$ results from both
the OM08 and the current work suggest that if the Euclidean nature of
the counts continue down to the natural confusion as an extrapolation
of our log N - log S, this suggests that the cutoff must be fairly
sharp below 1 $\mu$Jy.

\section{Conclusions}

We have presented a deep, new, high-resolution radio image of the
GOODS-N field and a catalog containing 1,230 secure radio sources in
the central 40\arcmin\,$\times$\,40\arcmin\ region.  The image reaches
an r.m.s.\ of $\sim$3.9\,$\mu$Jy\,beam$^{-1}$ in the best central
region. It is thus one of the deepest 1.4-GHz radio surveys ever
undertaken, well matched to the excellent {\em Spitzer}, {\em Hubble
Space Telescope} and {\em Chandra} data in GOODS-N, and to the
upcoming {\em Herschel}, SCUBA-2, and the Large Millimeter Telescope
imaging at 100--850\,$\mu$m and  $\sim$1.1-4mm, respectively, providing
an unbiased probe of star formation out to $z\sim3$ and a means to
test the calibration of other star-formation rate indicators.

Our analysis of the source size distribution and the counts provide
further evidence that the faint 1.4-GHz emitters have a median angular
size of $\sim$1.2\arcsec, that log $N$--log $S$ remains flat below
100\,$\mu$Jy, and that the natural confusion limit at 1.4\,GHz must be
near 1\,$\mu$Jy -- a prediction that can be tested in the coming years
using the Expanded VLA (EVLA) and e-MERLIN.

\acknowledgments

G.E.M.\ acknowledges financial support from NRAO and travel support
from the UK's Science and Technology Facilities Council via the UK
Astronomy Technology Centre at the Royal Observatory, Edinburgh. The
National Radio Astronomy Observatory is a facility of the National
Science Foundation operated under cooperative agreement by Associated
Universities, Inc

G.E.M.\ also acknowledges financial support for this work, part of the
Space Infrared Telescope Facility Legacy Science Program, was provided
by NASA through contract 1224666 issued by the Jet Propulsion
Laboratory, California Institute of Technology, under NASA contract
1407. We would also like to thank the anonymous referee for useful
suggestions to enhance the quality of the paper.

\clearpage

\begin{figure}[p]
   \epsscale{2.0}\plottwo{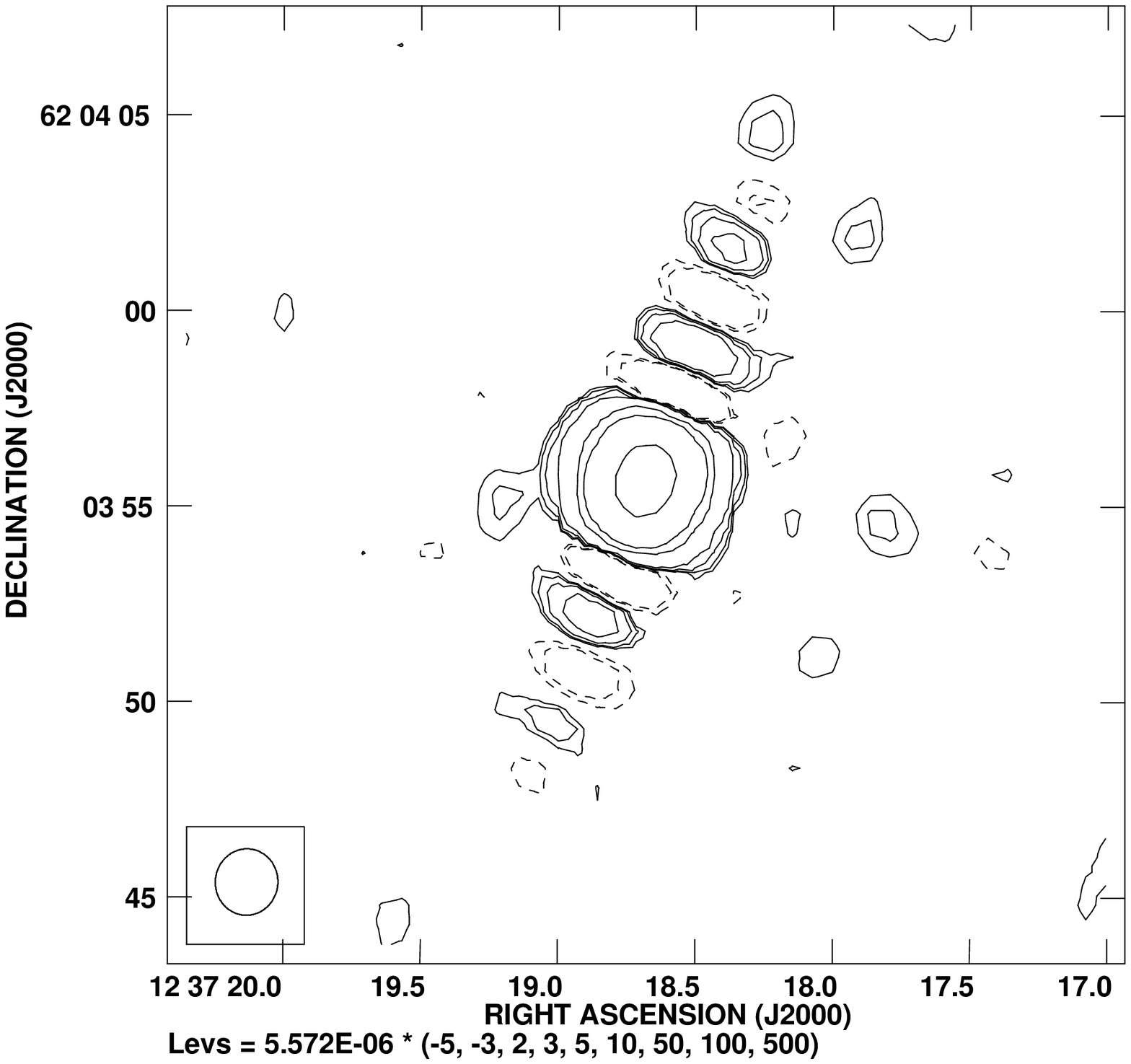}{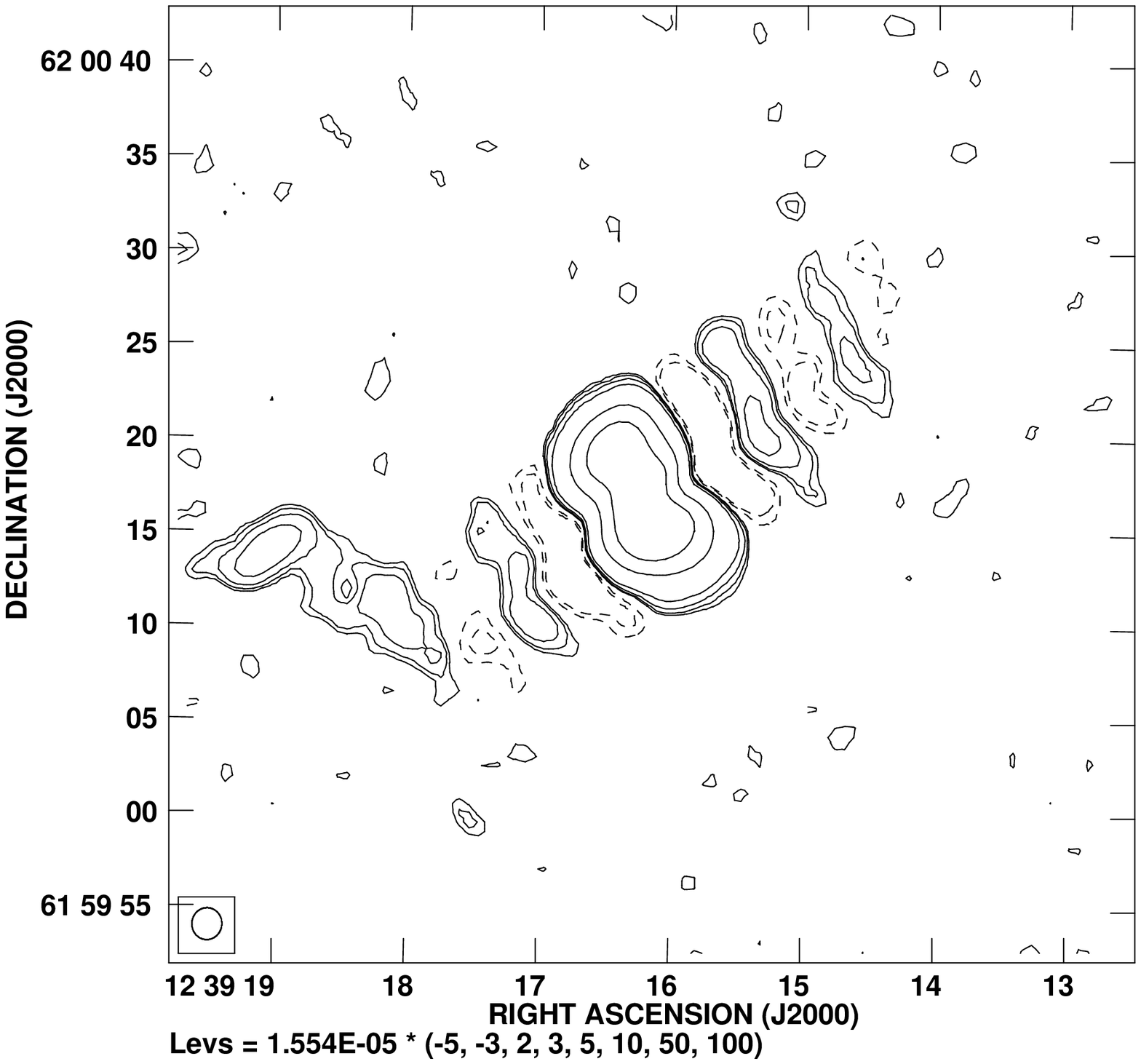} \caption{Extended
   sources in the GOODS-N field. Contour levels are listed on the
   figure and the synthesized beam (FWHM) is illustrated at the bottom
   left.  Several of the sources displayed here might be extended due
   to band-width smearing issues. The quasi-periodic artifacts are due
   to uncorrectable correlator errors. In practice, this pattern is
   not the same for every channel and thus cannot properly be remove
   from the images by CLEAN deconvolution.\label{kntr2}}
\end{figure}

\begin{figure}[p]
   \epsscale{2.0}\plottwo{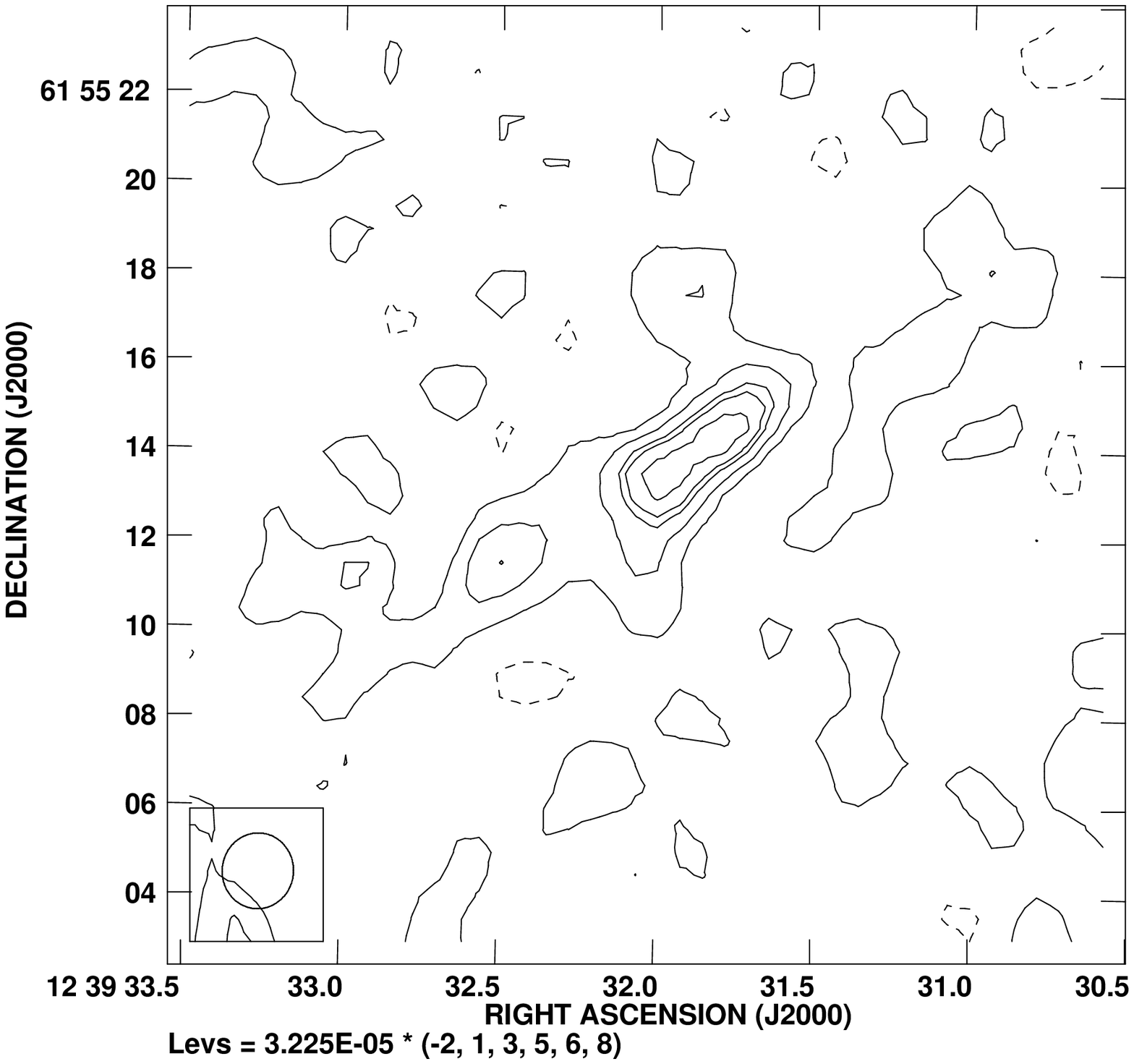}{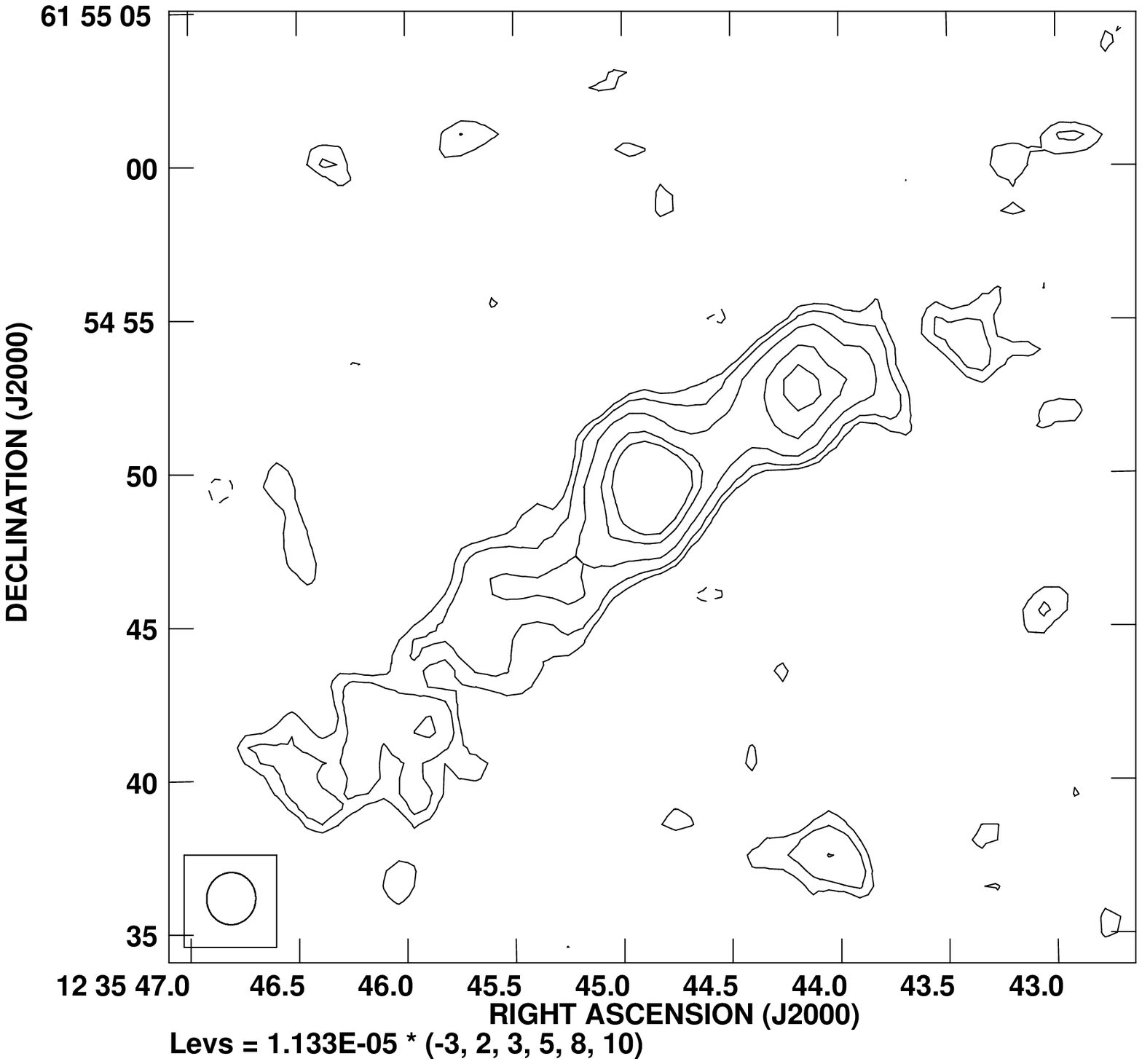}
   \caption{Cont...}
\end{figure}

\begin{figure}[p]
   \epsscale{2.0}\plottwo{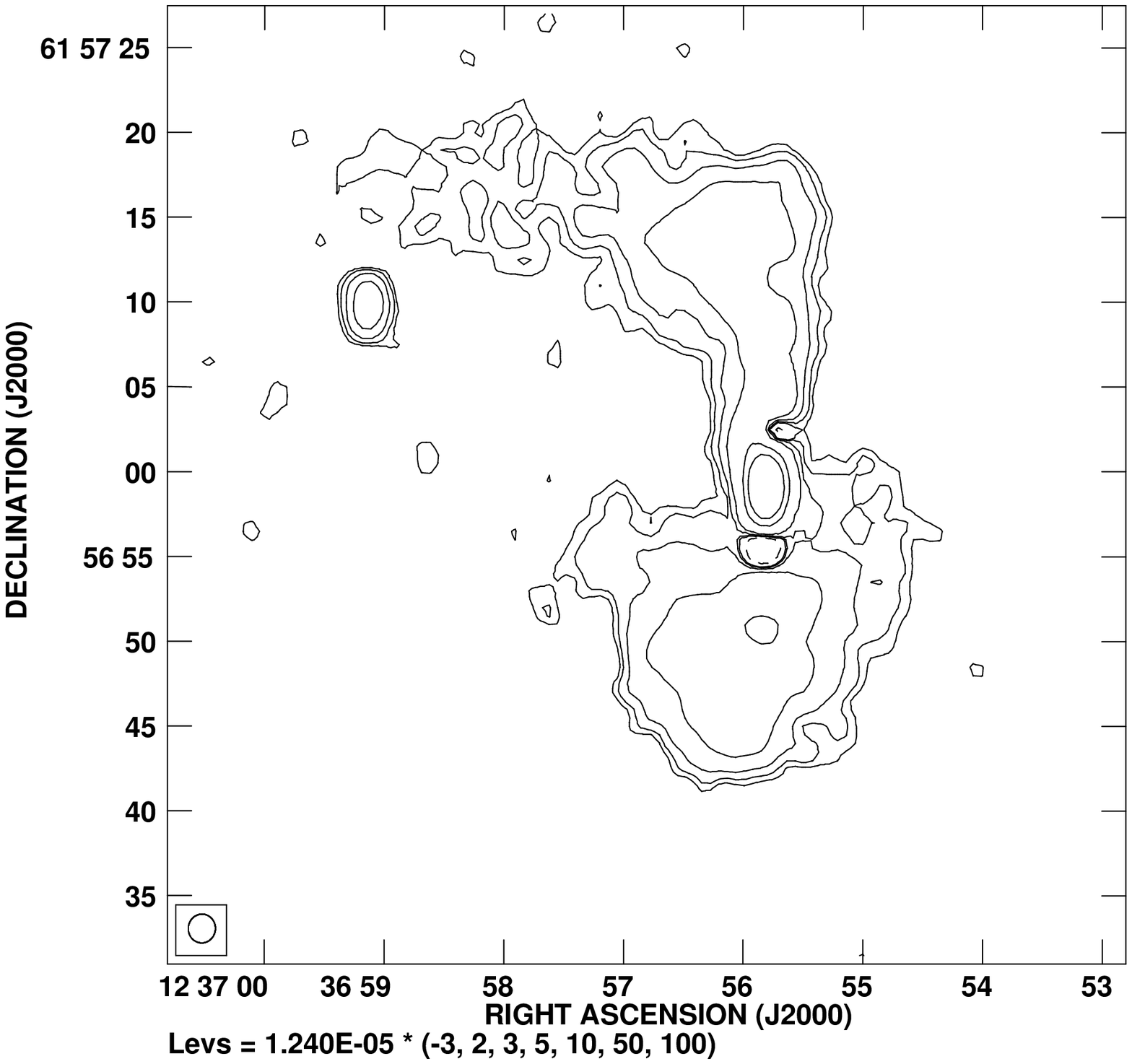}{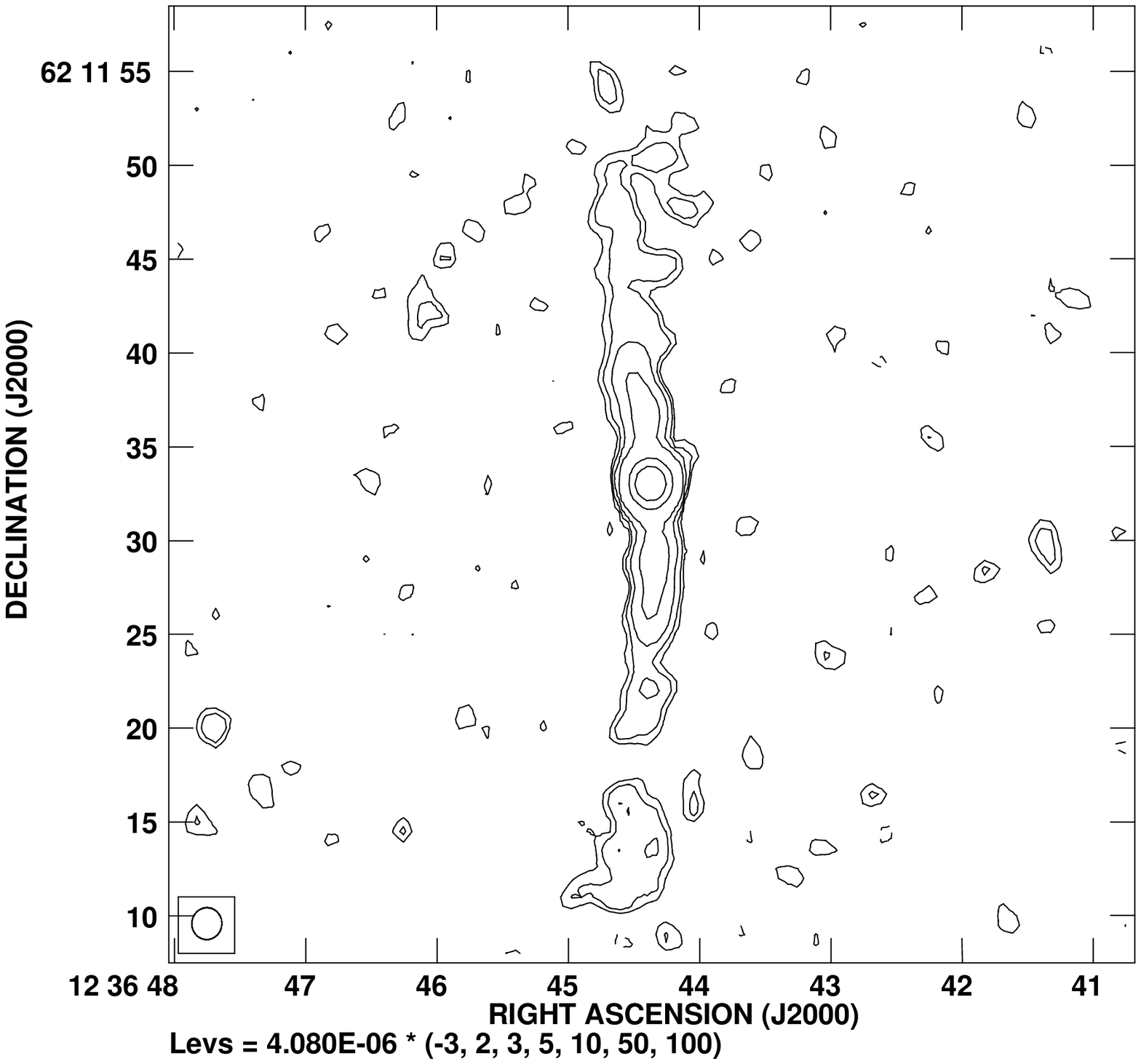}
   \caption{Cont...}
\end{figure}

\begin{figure}[p]
   \epsscale{2.0}\plottwo{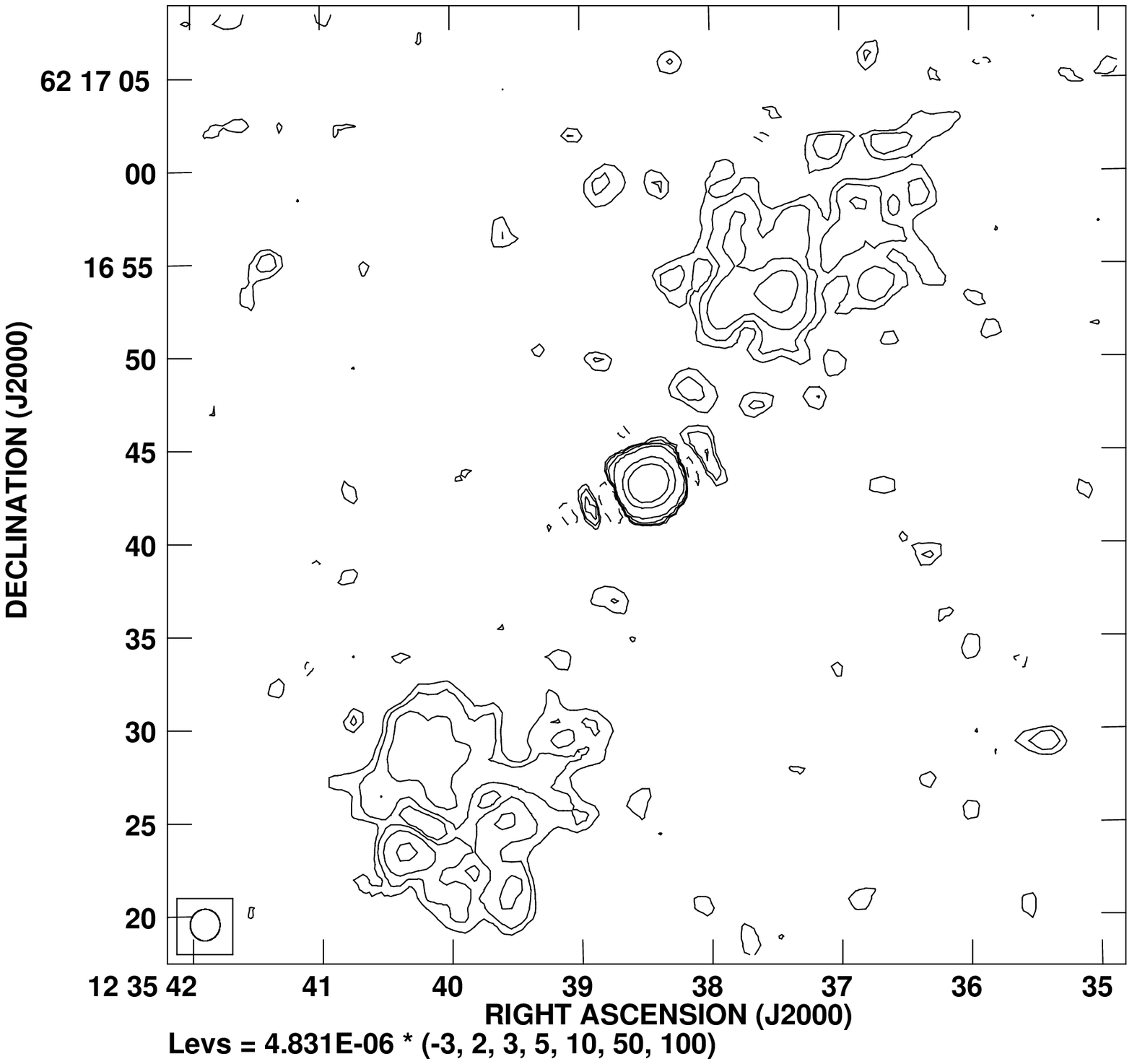}{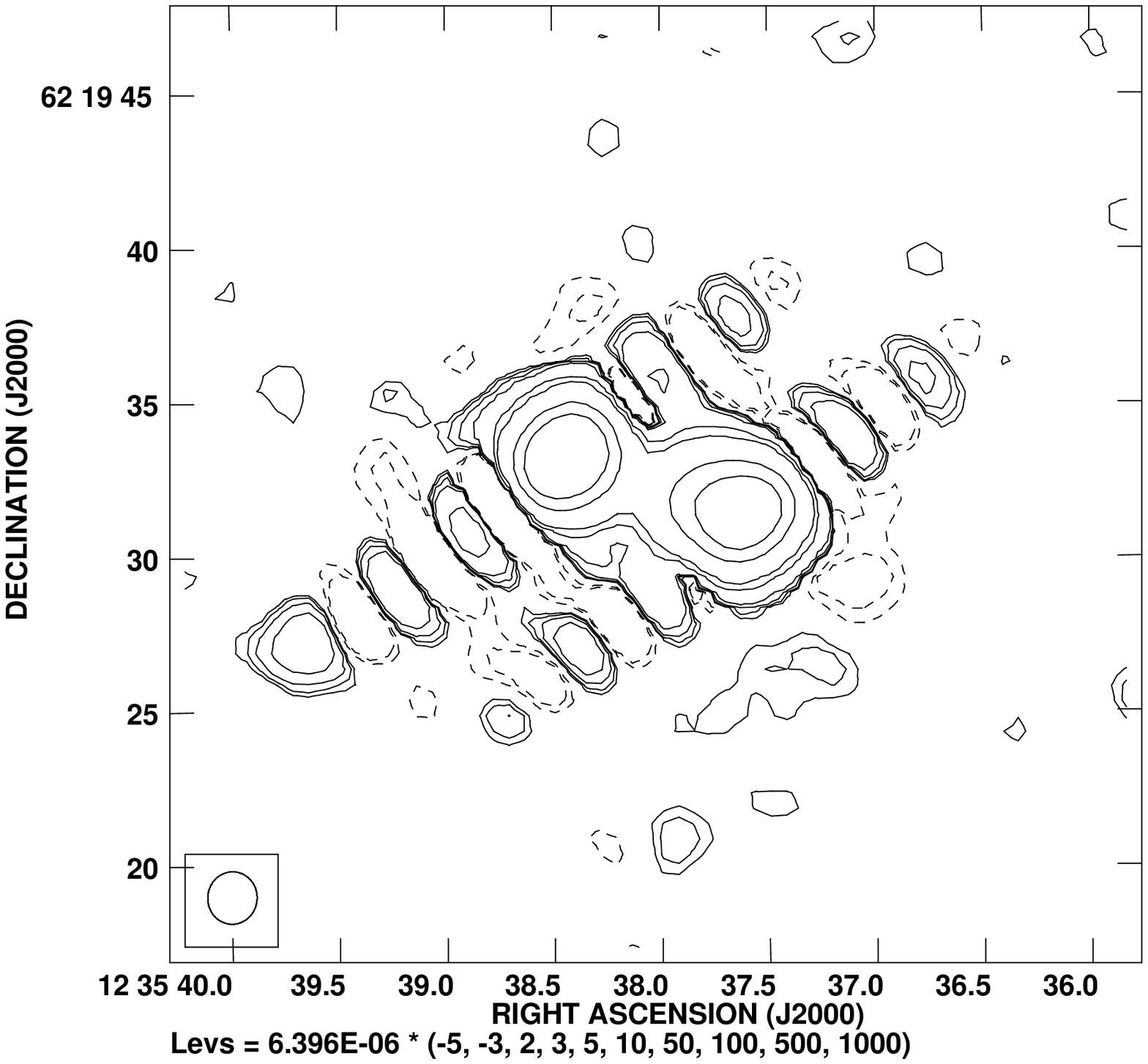}
   \caption{Cont...}
\end{figure}

\begin{figure}[p]
   \epsscale{2.0}\plottwo{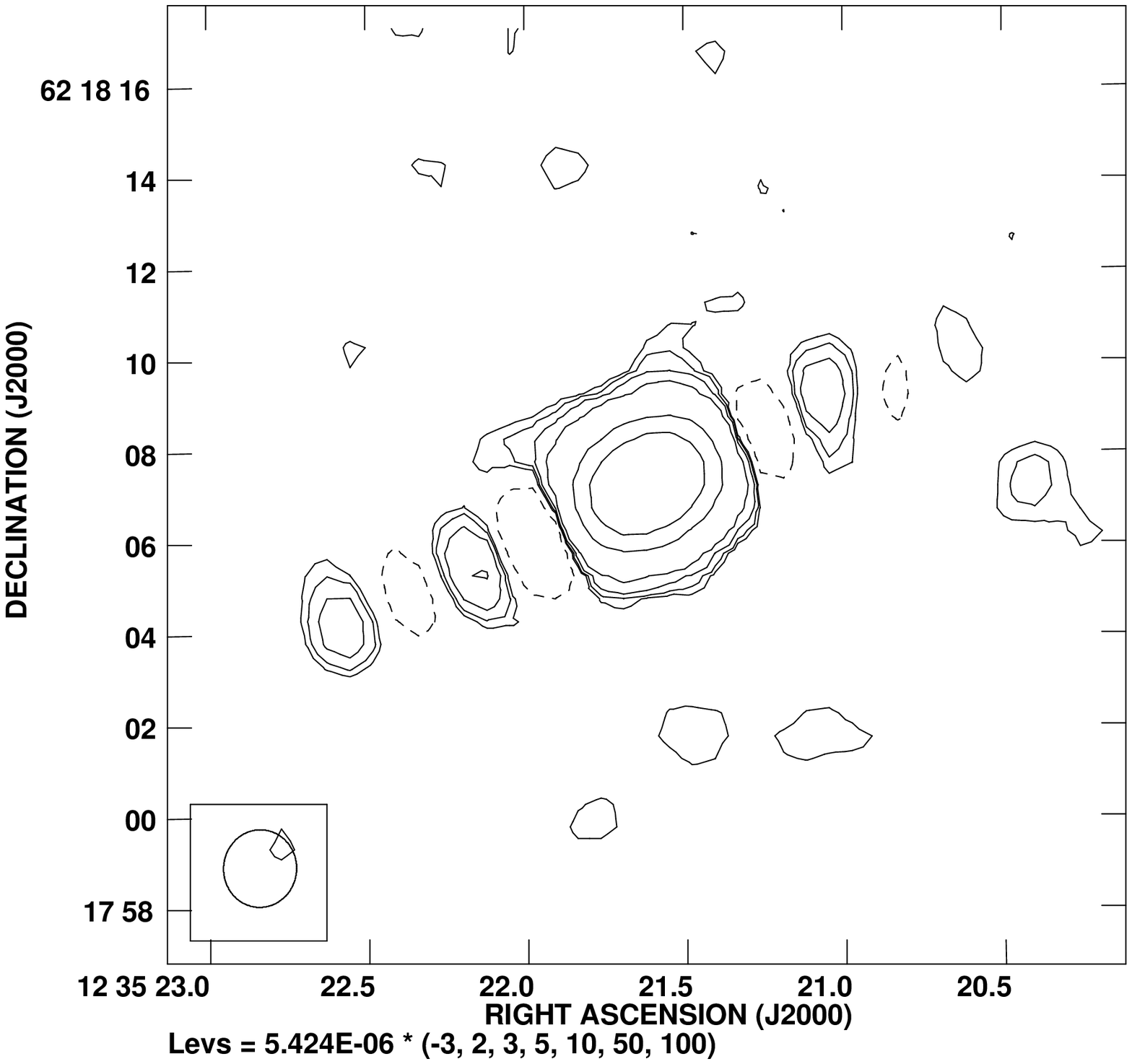}{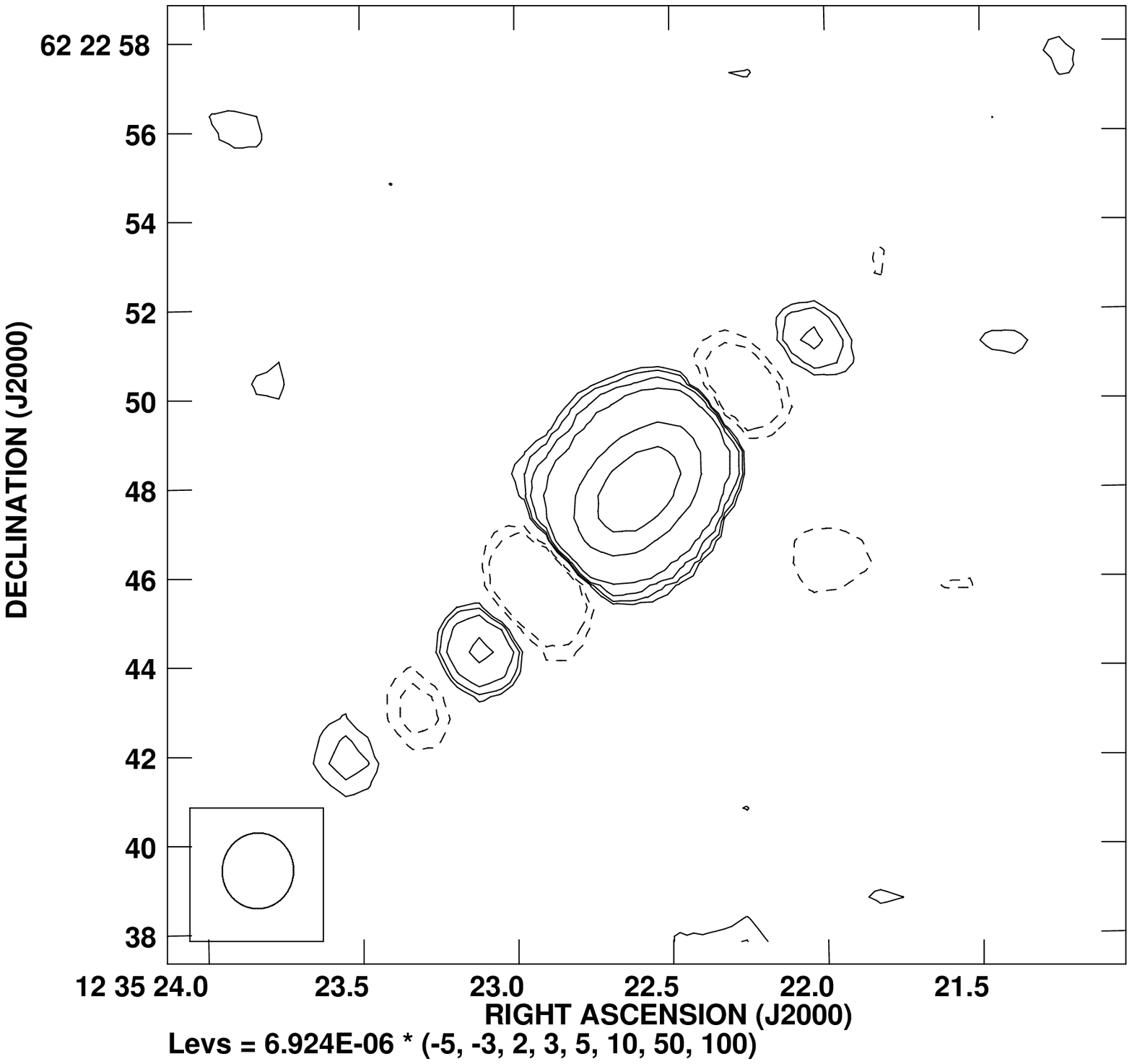}
   \caption{Cont...}
\end{figure}

\begin{figure}[p]
   \epsscale{2.0}\plottwo{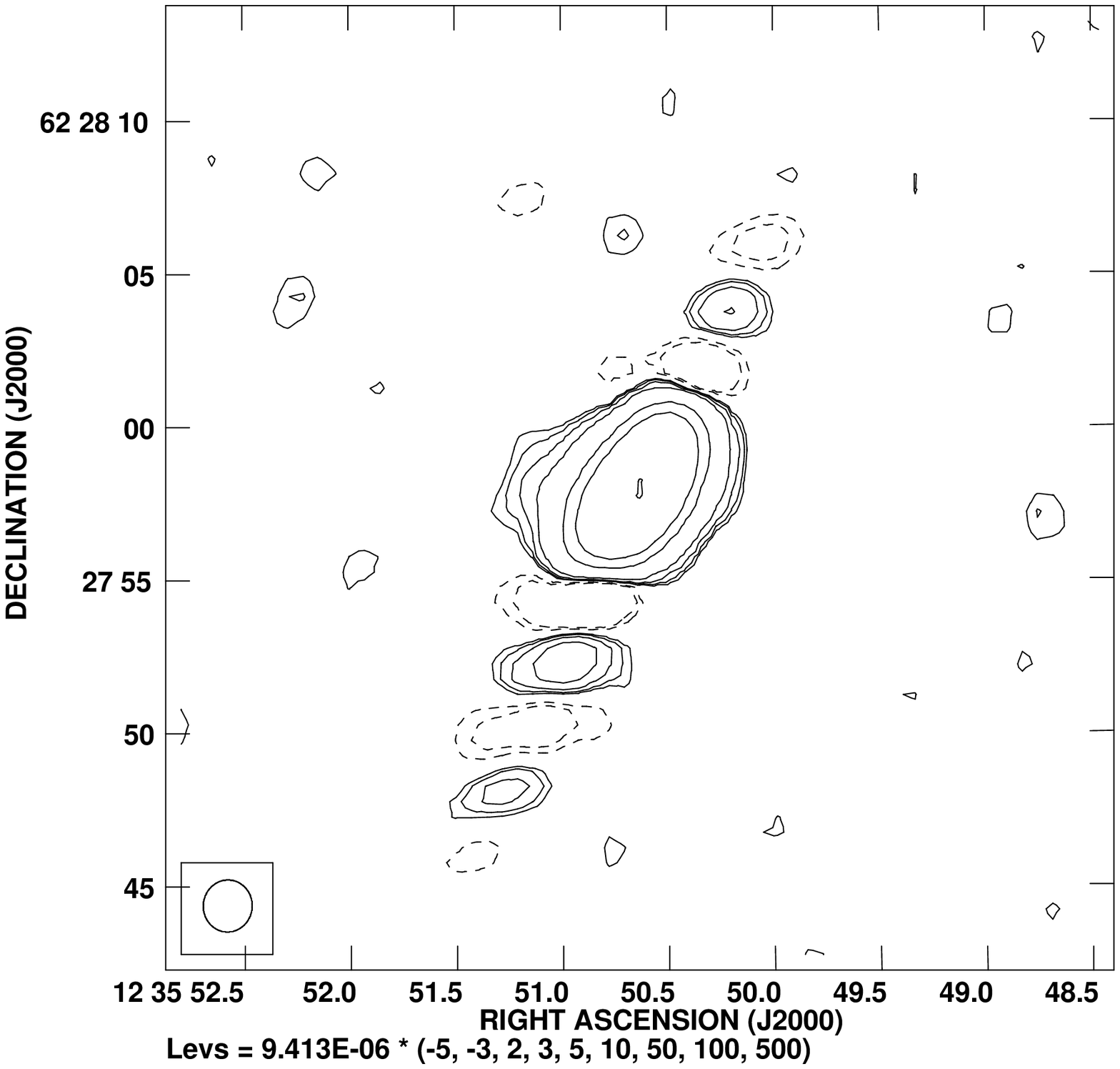}{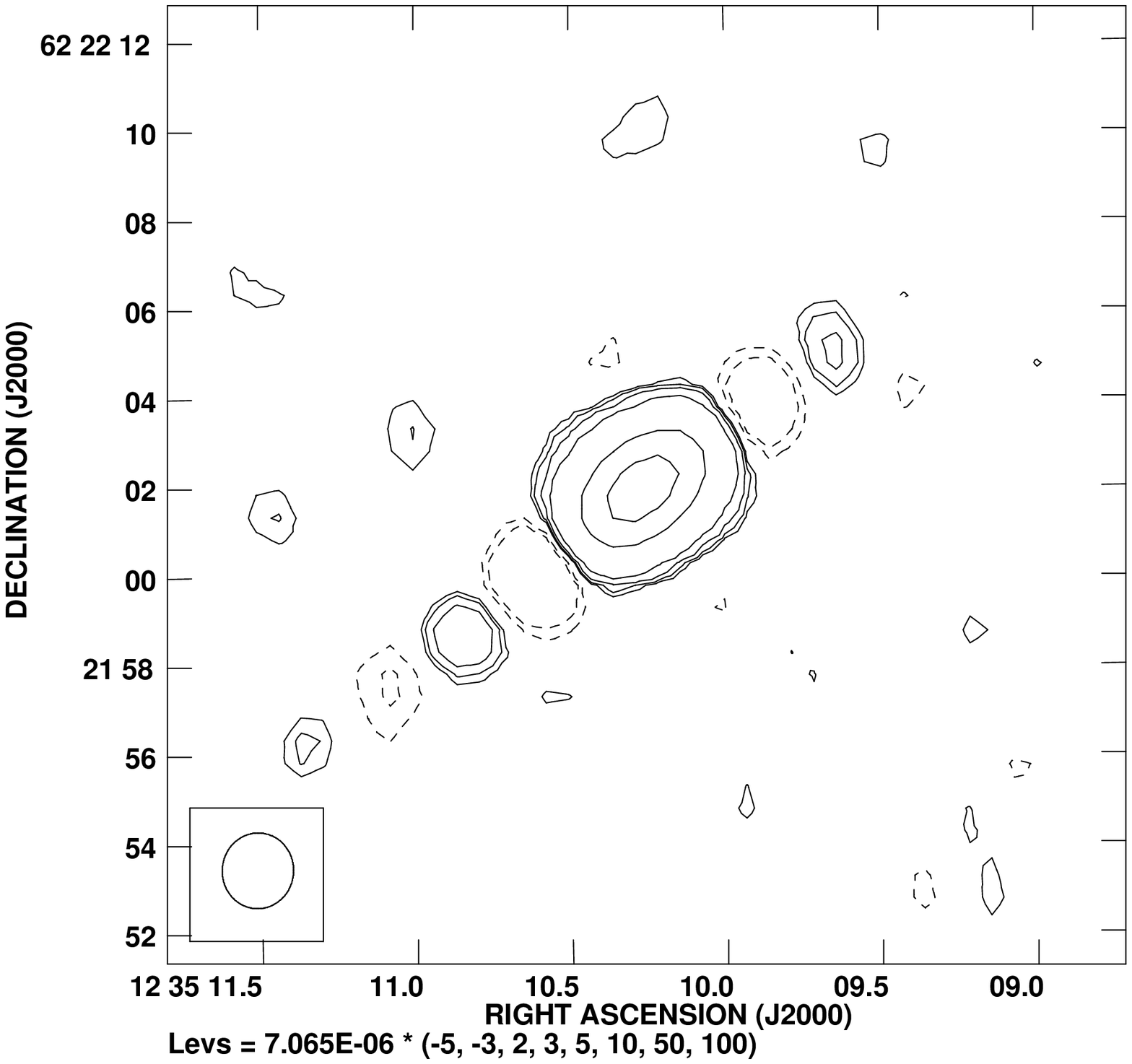}
   \caption{Cont...}
\end{figure}

\begin{figure}[p]
   \epsscale{2.0}\plottwo{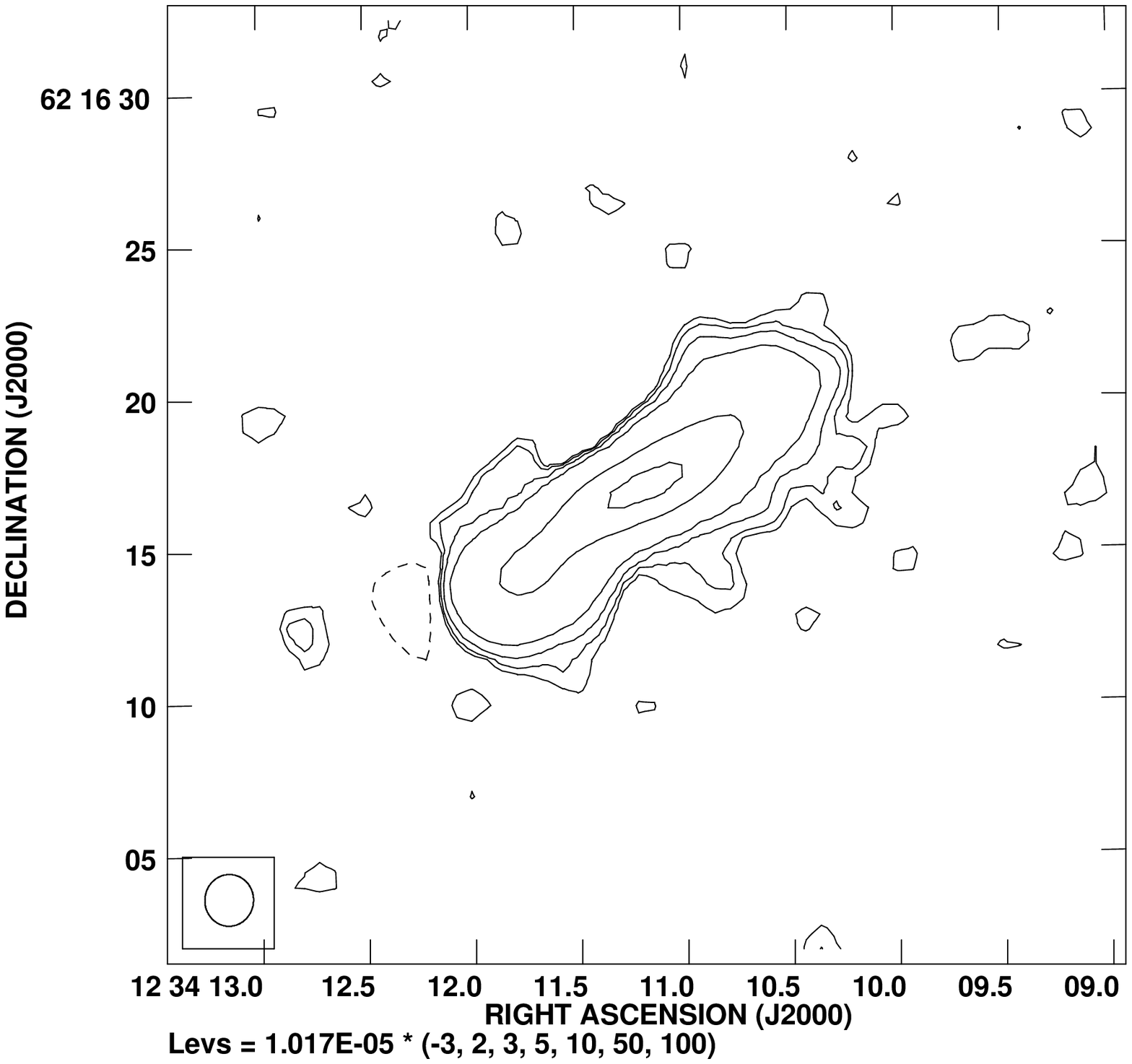}{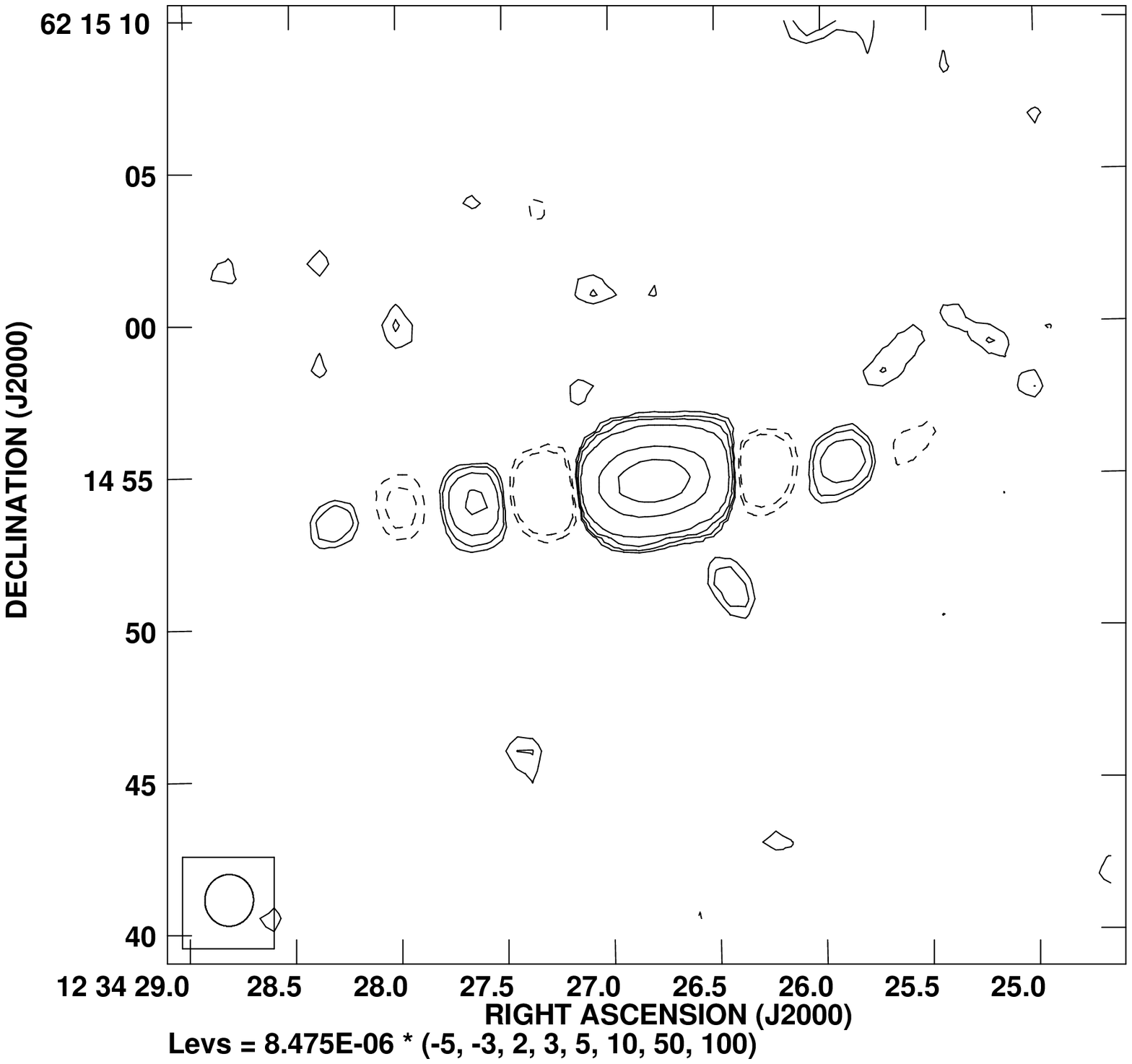}
   \caption{Cont...}
\end{figure}

\begin{figure}[p]
   \epsscale{2.0}\plottwo{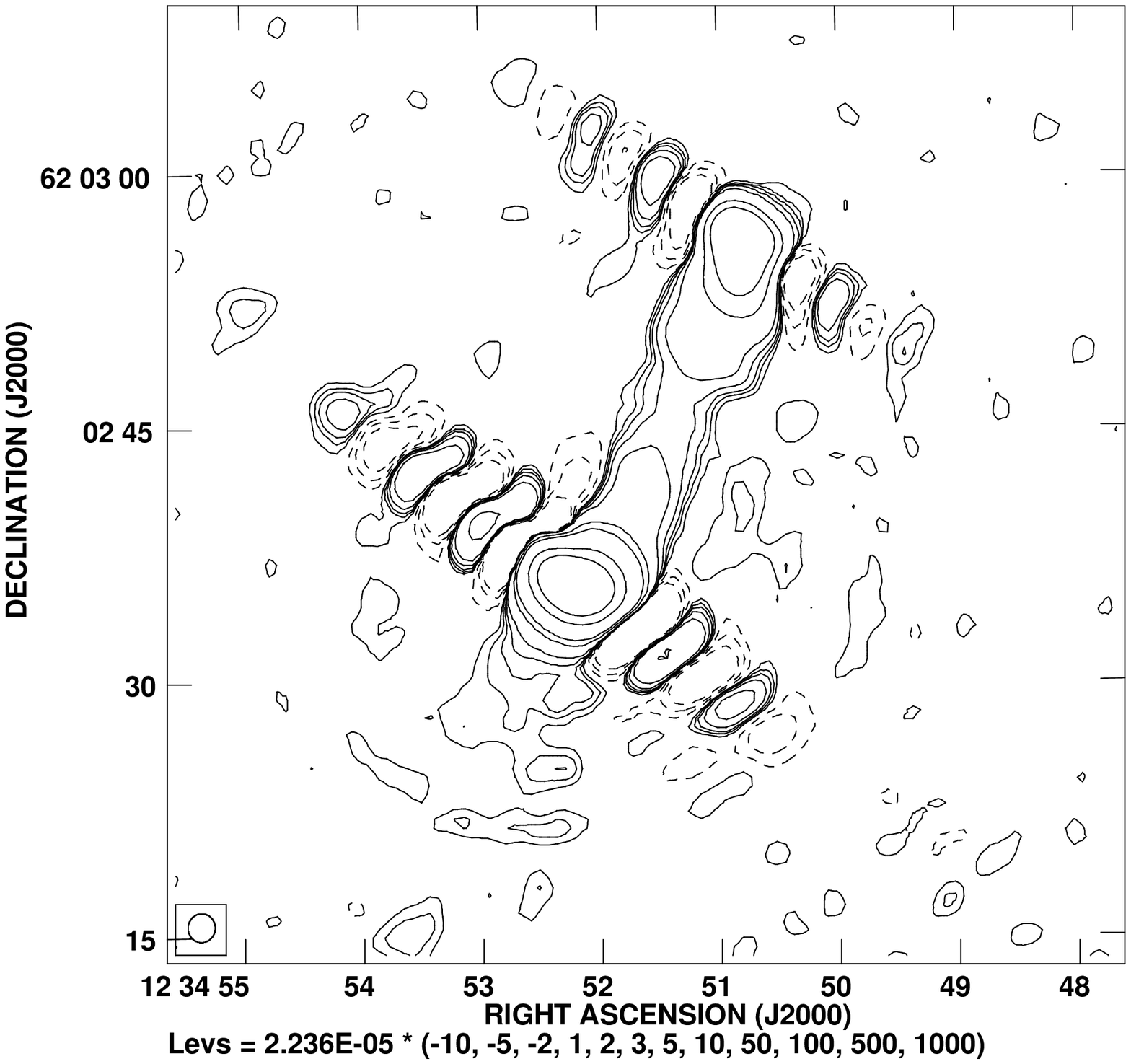}{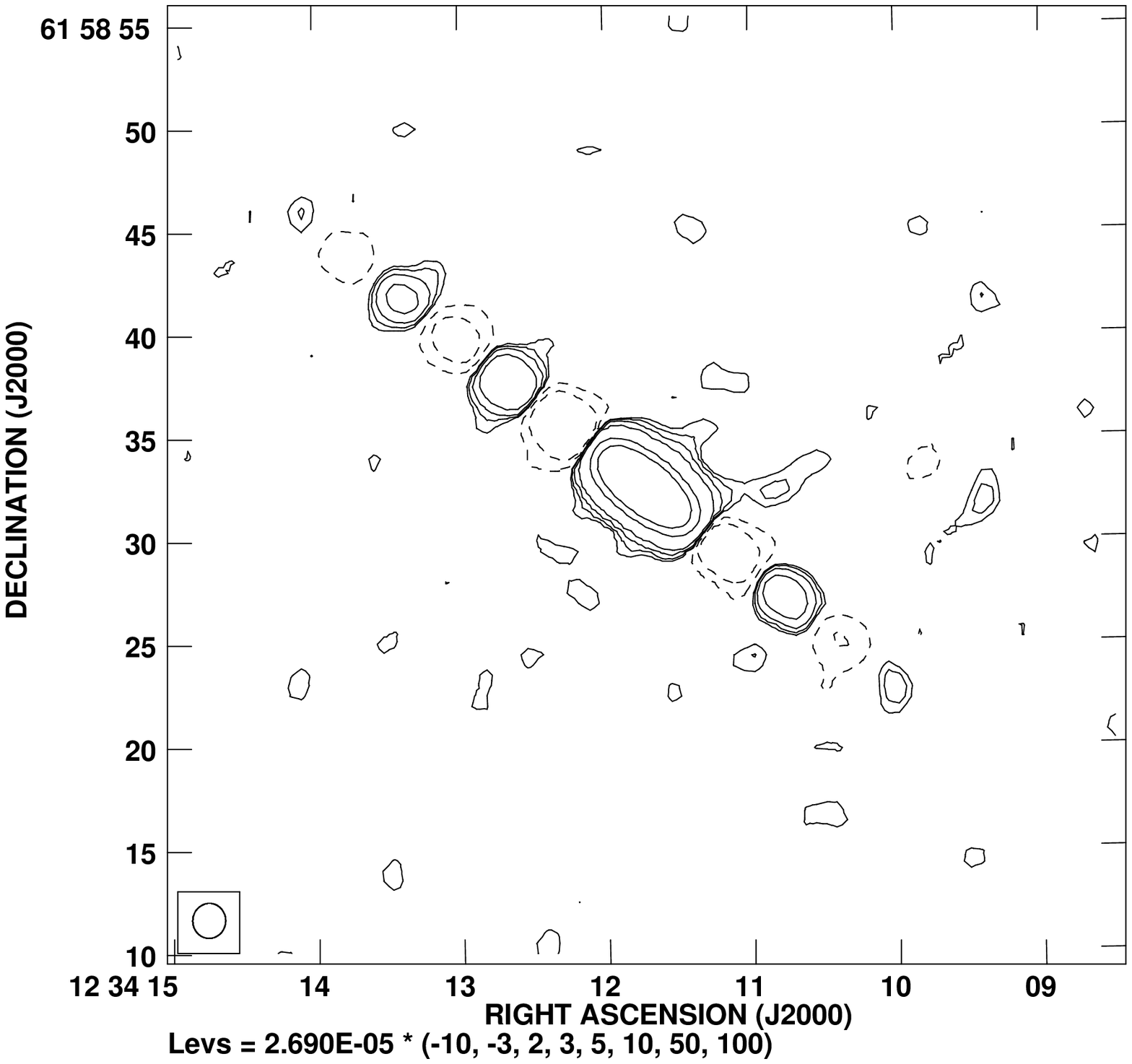}
   \caption{Brightest source in the HDF is in the upper
   right. Undeconvolvable sidelobes around this source are clearly
   visible and point to the phase center.}\label{kntr3}
\end{figure}

\begin{figure}[p]
   \epsscale{2.0}\plotone{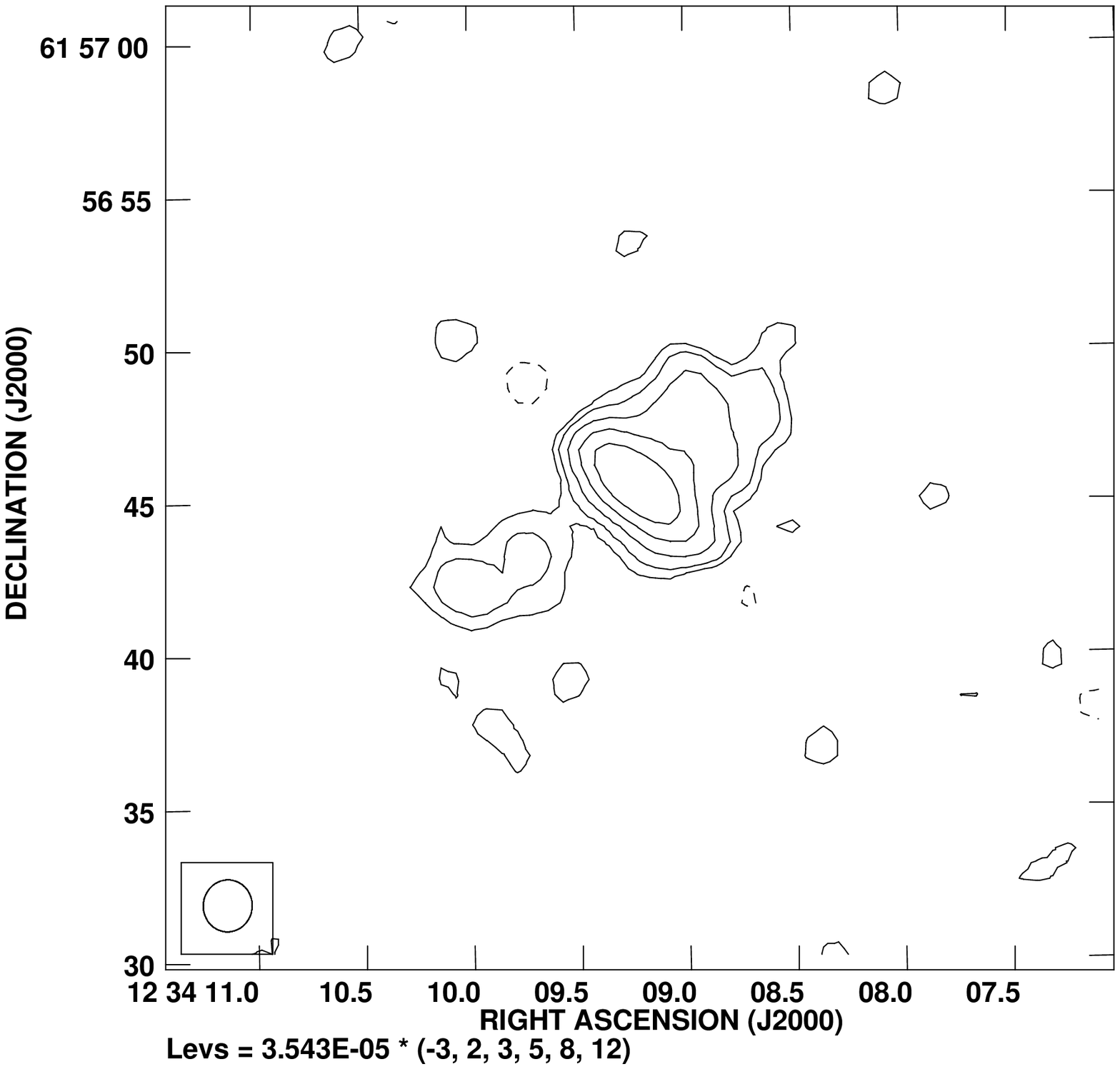}
   \caption{Cont...}\label{kntr4}
\end{figure}

\clearpage


\end{document}